# A Review of Theory and Practice in Scientometrics[1]

**John Mingers**
Kent Business School, University of Kent, Canterbury CT7 2PE, UK
j.mingers@kent.ac.uk
01227 824008

**Loet Leydesdorff**
Amsterdam School of Communication Research (ASCoR), University of Amsterdam, PO Box 15793, 1001 NG Amsterdam, The Netherlands
loet@leydesdorff.net

**Abstract**

Scientometrics is the study of the quantitative aspects of the process of science as a communication system. It is centrally, but not only, concerned with the analysis of citations in the academic literature. In recent years it has come to play a major role in the measurement and evaluation of research performance. In this review we consider: the historical development of scientometrics, sources of citation data, citation metrics and the "laws" of scientometrics, normalisation, journal impact factors and other journal metrics, visualising and mapping science, evaluation and policy, and future developments.

## 1. HISTORY AND DEVELOPMENT OF SCIENTOMETRICS

Scientometrics was first defined by Nalimov (1971, p. 2) as developing "the quantitative methods of the research on the development of science as an informational process". It can be considered as the study of the quantitative aspects of science and technology seen as a process of communication. Some of the main themes include ways of measuring research quality and impact, understanding the processes of citations, mapping scientific fields and the use of indicators in research policy and management. Scientometrics focuses on communication in the sciences, the social sciences, and the humanities among several related fields:

*Bibliometrics* – "The application of mathematics and statistical methods to books and other media of communication" (Pritchard, 1969, p. 349). This is the original area of study covering books and publications generally. The term "bibliometrics" was first proposed by Otlet (1934); see also Rousseau (2014).

[1] We would like to acknowledge helpful comments from Anne-Wil Harzing, David Pendlebury, Ronald Rousseau and an anonymous referee

*Scientometrics* – "The quantitative methods of the research on the development of science as an informational process" (Nalimov & Mulcjenko, 1971, p. 2). This field concentrates specifically on science (and the social sciences and humanities).

*Informetrics* – "The study of the application of mathematical methods to the objects of information science" (Nacke, 1979, p. 220). Perhaps the most general field covering all types of information regardless of form or origin (Bar-Ilan, 2008; Egghe & Rousseau, 1990; Egghe & Rousseau, 1988; Wilson, 1999).

*Webometrics* – "The study of the quantitative aspects of the construction and use of information resources, structures and technologies on the Web drawing on bibliometric and informetric approaches (Björneborn & Ingwersen, 2004, p. 1217; Thelwall & Vaughan, 2004; Thelwall et al., 2005). This field mainly concerns the analysis of web pages as if they were documents.

*Altmetrics* – "The study and use of scholarly impact measures based on activity in online tools and environments" (Priem, 2014, p. 266). Also called Scientometrics 2.0, this field replaces journal citations with impacts in social networking tools such as views, downloads, "likes", blogs, Twitter, Mendelay, CiteULike.

In this review we concentrate on scientometrics as that is the field most directly concerned with the exploration and evaluation of scientific research. In fact, traditionally these fields have concentrated on the observable or measurable aspects of communications – external borrowings of books rather than in-library usage; citations of papers rather than their reading – but currently online access and downloads provide new modes of usage and this leads to the developments in webometrics and altmetrics that will be discussed later. In this section we describe the history and development of scientometrics (de Bellis, 2014; Leydesdorff & Milojevic, 2015) and in the next sections explore the main research areas and issues.

Whilst scientometrics can, and to some extent does, study many other aspects of the dynamics of science and technology, in practice it has developed around one core notion – that of the citation. The act of citing another person's research provides the necessary linkages between people, ideas, journals and institutions to constitute an empirical field or network that can be analysed quantitatively. Furthermore, the citation also provides a linkage in time – between the previous publications of its references and the later appearance of its citations. This in turn stems largely from the work of one person – Eugene Garfield – who identified the importance of the citation and then promulgated the idea of the *Science Citation Index* (SCI) in the 1950's (and the company the *Institute for Scientific Information, ISI*, to maintain it) as a database for capturing citations (Garfield, 1955; Garfield, 1979)[2]. Its initial purpose was not research evaluation, but rather help for researchers to search the literature more effectively – citations could work well as index or search terms, and also enabled unfamiliar authors to be discovered. The *SCI* was soon joined by the *Social Sciences Citation Index* (SSCI, in 1973) and the *Arts & Humanities Citation Index* (A&HCI; since 1978), and eventually taken over by Thomson Corporation who converted it into the *Web of Science* as part of their *Web of Knowledge* platform[3]. In 2013, the *SCI* covered 8,539 journals, the *SSCI* 3,080 journals, and the *A&HCI* approximately 1,700 journals.

[2] It was first realised in 1964

The SCI was soon recognized as having great value for the empirical study of the practice of science. The historian, Derek de Solla Price (1963; 1965), was one of the first to see the importance of networks of papers and authors and also began to analyse scientometric processes, leading to the idea of cumulative advantage (Price, 1976), a version of "success to the successful" (Senge, 1990) or "success breeds success (SBS)" also known as the Matthew[4] effect (Merton, 1968; Merton, 1988). Price identified some of the key problems that would be addressed by scientometricians: mapping the "invisible colleges" (Crane, 1972) informally linking highly cited researchers at the research frontiers (cf co-authorship networks and co-citation analysis (Marshakova, 1973; Small, 1973)): studying the links between productivity and quality in that the most productive are often the most highly cited (cf the h-index); and investigating citation practices in different fields (cf normalization). In 1978, Robert K. Merton, a major sociologist, was one of the editors of a volume called *Towards a Metric of Science: The Advent of Science Indicators* (Elkana et al., 1978) which explored many of these new approaches. Scientometrics was also developing as a discipline with the advent of the journal *Scientometrics* in 1978; a research unit in the Hungarian Academy of Sciences and scientific conferences and associations.

At the same time as scientometrics research programmes were beginning, the first links to research evaluation and the use of citation analysis in policy making also occurred. For example, the ISI data was included in the (US) National Science Board's *Science Indicators Reports* in 1972 and was used by the OECD. Garfield and Sher (Garfield & Sher, 1963) developed a measure for evaluating journals – the *impact factor (IF)* – that has been for many years a standard despite its many flaws. Journals with this specific policy focus appeared such as *Research Policy, R&D Management* and *Research Evaluation.*

During the 1990s and 2000s several developments have occurred. The availability and coverage of the citation databases has increased immensely. The WoS itself includes many more journals and also conferences proceedings, although its coverage in the social sciences and humanities is still limited. It also does not yet cover books adequately although there are moves in that direction. A rival, *Scopus,* has also appeared form the publishers Elsevier. However, the most interesting challenger is *Google*

[3] Web of Knowledge has now reverted to Web of Science.
[4] Named after St Matthew (25:29): "For unto everyone that hath shall be given .. from him that hath not shall be taken away"

*Scholar* which works in an entirely different way – searching the web rather than collecting data directly. Whilst this extension of coverage is valuable, it also leads to problems of comparison with quite different results appearing depending on the databases used.

Secondly, a whole new range of metrics has appeared superseding, in some ways, the original ones such as total number of citations and citations per paper (cpp). The h-index (Costas & Bordons, 2007; Egghe, 2010; Glänzel, 2006; Hirsch, 2005; Mingers, 2008b; Mingers et al., 2012) is one that has become particularly prominent, now available automatically in the databases. It is transparent and robust but there are many criticisms of its biases. In terms of journal evaluation, several new metrics have been developed such as SNIP (Moed, 2010b) and SCImago Journal Rank (SJR) (González-Pereira et al., 2010; Guerrero-Bote & Moya-Anegón, 2012) which aim to take into account the differential citation behaviours of different disciplines, e.g., some areas of science such as biomedicine cite very highly and have many authors per paper; other areas, particularly some of the social sciences, mathematics and the humanities do not cite so highly.

A third, technical, development has been in the mapping and visualization of bibliometric networks. This idea was also initiated by Garfield who developed the concept of "historiographs" (Garfield et al., 1964), maps of connections between key papers, to reconstruct the intellectual forebears of an important discovery. This was followed by co-citation analysis which used multivariate techniques such as factor analysis, multi-dimensional scaling and cluster analysis to analyse and map the networks of highly related papers which pointed the way to identifying research domains and frontiers (Marshakova, 1973; Small, 1973). And also co-word analysis that looked at word pairs from titles, abstracts or keywords and drew on the actor network theory of Callon and Latour (Callon et al., 1983). New algorithms and mapping techniques such as the Blondel algorithm (Blondel et al., 2008) and the Pajek mapping software have greatly enhanced the visualization of high-dimensional datasets (de Nooy et al., 2011).

But perhaps the most significant change, which has taken scientometrics from relative obscurity as a statistical branch of information science to playing a major, and often much criticised, role within the social and political processes of the academic community, is the drive of governments and official bodies to monitor, record and evaluate research performance. This itself is an effect of the neo-liberal agenda of "new public management" (NPM) (Ferlie et al., 1996) and its requirements of transparency and accountability. This occurs at multiple levels – individuals, departments and research groups, institutions and, of course, journals – and has significant consequences in terms of jobs and promotion, research grants, and league tables. In the past, to the extent that this occurred it did so through a process of peer review with the obvious drawbacks of subjectivity, favouritism and conservatism (Bornmann, 2011; Irvine et al., 1985). But now, partly on cost grounds, scientometrics are being called into play and the rather ironic result is that instead of merely reflecting or mapping a pre-given reality, scientometrics methods are actually shaping that reality through their performative effects on academics and researchers (Wouters, 2014).

At the same time, the discipline of science studies itself has bi- (or tri-) furcated into at least three elements – the quantitative study of science indicators and their behaviour, analysis and metrication from a positivist perspective. A more qualitative, sociology-of-science, approach that studies the social and political processes lying behind the generation and effects of citations, generally from a constructivist perspective. And a third stream of research that is interested in policy implications and draws on both the other two.

Finally, in this brief overview, we must mention the advent of the Web and social networking. This has brought in the possibility of alternatives to citations as ways of measuring impact (if not quality) such as downloads, views, "tweets", "likes", and mentions in blogs. Together, these are known as "altmetrics" (Priem, 2014), and whilst they are currently underdeveloped, they may well come to rival citations in the future. There are also academic social networking sites such as *ResearchGate (www.researchgate.net), CiteULike (citeulike.org), academia.edu (www.academia.edu), RePEc (repec.org)* and *Mendeley (www.mendeley.com)* which in some cases have their own research metrics. *Google Scholar* can produces profiles of researchers, including their h-index, and *Publish or Perish* (Harzing, 2007) enhances searches of *Scholar* with the Harzing website (www.harzing.com) being a repository for multiple journals ranking lists in the field of business and management.

## 2. SOURCES OF CITATIONS DATA

Clearly for the quantitative analysis of citations to be successful, there must be comprehensive and accurate sources of citation data. The major source of citations in the past was the Thomson Reuters ISI *Web of Science* (WoS) which is a specialised database covering all the papers in around 12,000 journals[5] . It also covers conference proceedings[6] and is beginning to cover books[7]. Since 2004, a very similar rival database is available from Elsevier called *Scopus*[8] that covers 20,000 journals and also conferences and books. *Scopus* retrieves back until 1996, while WoS is available for all years since 1900. These two databases have been the traditional source for most major scientometrics exercises, for example by the *Centre for Science and Technology Studies* (CWTS) which has specialised access to them. More recently (2004), an alternative source has been provided by *Google Scholar* (GS). This works in an entirely different way, by searching the Web for documents that have references to papers and books rather than inputting data from journals. It is best accessed through a software program called *Publish or Perish*[9]. Both of these resources are free whilst access to *WoS* and *Scopus* are subscription-based and offer different levels of accessibility depending on the amount of payment thus leading to differential access for researchers.

[5] http://wokinfo.com/essays/journal-selection-process/
[6] http://wokinfo.com/products_tools/multidisciplinary/webofscience/cpci/cpciessay/
[7] http://thomsonreuters.com/book-citation-index/
[8] http://www.elsevier.com/online-tools/scopus/content-overview

Many studies have shown that the coverage of WoS and Scopus differs significantly between different fields, particularly between the natural sciences, where coverage is very good, the social sciences where it is moderate and variable, and the arts and humanities where it is generally poor (HEFCE, 2008; Larivière et al., 2006; Mahdi et al., 2008; Moed & Visser, 2008)[10],. In contrast, the coverage of GS is generally higher, and does not differ so much between subject areas, but the reliability and quality of its data can be poor (Amara & Landry, 2012).

Van Leeuwen (2006), in a study of Delft University between 1991 and 2001 found that in fields such as architecture and technology, policy and management the proportion of publication in WoS and the proportion of references to ISI material was under 30% while for applied science it was between 70% and 80%. Across the social sciences, the proportions varied between 20% for political science and 50% for psychology. Mahdi et al. (2008) studied the results of the 2001 RAE in the UK and found that while 89% of the outputs in biomedicine were in WoS, the figures for social science and arts & humanities were 35% and 13% respectively. CWTS (Moed et al., 2008) was commissioned to analyse the 2001 RAE and found that the proportions of outputs contained in WoS and Scopus respectively were: Economics (66%, 72%), Business and Management (38%, 46%), Library and Information Science (32%, 34%) and Accounting and Finance (22%, 35%).

There are several reasons for the differential coverage in these databases (Archambault et al., 2006; Larivière et al., 2006; Nederhof, 2006) and we should also note that the problem is not just the publications that are not included, but also that the publications that are included have lower citations recorded since many of the citing sources are not themselves included. The first reason is that in science almost all research publications appear in journal papers (which are largely included in the databases), but in the social sciences and even more so in humanities books are seen as the major form of research output. Secondly, there is a greater prevalence of the "lone scholar" as opposed to the team approach that is necessary in the experimental sciences and which results in a greater number of publications (and hence citations) overall. As an extreme example, a paper in *Physics Letters B* (Aad et al., 2012) in 2012 announcing the discovery of the Higgs Boson has 2,932 authors and already has over 4000 citations. These outliers can distort bibliometrics analyses as we shall see (Cronin, 2001). Thirdly, a significant number of social science and humanities journals are not, or have not chosen to become, included in WoS, the accounting and finance field being a prime example. Finally, in social science and humanities a greater proportion of publications are directed at the general public or specialised constituencies such as practitioners and these "trade" publications or reports are not included in the databases.

---

[9] http://www.harzing.com/pop.htm
[10] Higher Education Funding Council for England

There have also been many comparisons of WoS, Scopus and Google Scholar across a range of disciplines (Adriaanse & Rensleigh, 2013; Amara & Landry, 2012; Franceschet, 2010; García-Pérez, 2010; Harzing & van der Wal, 2008; Jacso, 2005; Meho & Rogers, 2008; Meho & Yang, 2007). The general conclusions of these studies are:

- That the coverage of research outputs, including books and reports, is much higher in GS, usually around 90%, and that this is reasonably constant across the subjects. This means that GS has a comparatively greater advantage in the non-science subjects where Scopus and WoS are weak.
- Partly, but not wholly, because of the coverage, GS generates a significantly greater number of citations for any particular work. This can range from two times to five times as many. This is because the citations come from a wide range of sources, not being limited to the journals that are included in the other databases.
- However, the data quality in GS is very poor with many entries being duplicated because of small differences in spellings or dates and many of the citations coming from a variety of non-research sources. With regard to the last point, it could be argued that the type of citation does not necessarily matter – it is still impact.

Typical of these comparisons is Mingers and Lipitakis (2010) who reviewed all the publications of three UK business schools from 1980 to 2008. Of the 4,600 publications in total, 3,023 were found in GS, but only 1,004 in WoS. None of the books, book chapters, conference papers or working papers were in WoS[11]. In terms of number of citations, the overall mean cites per paper (cpp) in GS was 14.7 but only 8.4 in WoS. It was also found that these rates varied considerably between fields in business and management, a topic to be taken up in the section on normalization. When taken down to the level of individual researchers the variation was even more noticeable both in terms of the proportion of outputs in WoS and the average number of citations. For example, the most prolific researcher had 109 publications. 92% were in GS, but only 40% were in WoS. The cpp in GS was 31.5, but in WoS it was 12.3. Generally, where papers were included in both sources GS cites were around three times greater.

With regard to data quality, Garcia-Perez (2010) studied papers of psychologists in WoS, GS, and PsycINFO[12]. GS recorded more publications and citations than either of the other sources, but also had a large proportion of incorrect citations (16.5%) in comparison with 1% or less in the other

[11] Most studies do not include WoS for books, which is still developing (Leydesdorff & Felt, 2012).
[12] PsycINFO is an abstracting and indexing database of the American Psychological Association with more than 3 million records devoted to peer-reviewed literature in the behavioural sciences and mental health

sources. The errors included not supplying usable links to the citation, phantom citations, duplicate links pointing to the same citing paper, or reprints published in different sources. Adriaanse and Rensleigh (2013) studied environmental scientists in WoS, Scopus and GS and made a comprehensive record of the inconsistencies that occurred in all three across all bibliometric record fields - data export, author, article title, page numbers, references and document type. There were clear differences with GS having 14.0% inconsistencies, WoS 5.4%, and Scopus only 0.4%. Similar problems with GS were also found by Jacso (2008) and Harzing and van der Wal (2008).

To summarise this section, there is general agreement at this point in time that bibliometric data from WoS or Scopus is adequate to conduct research evaluations in the natural and formal sciences where the coverage of publications is high, but it is not adequate in the social sciences or humanities, although, of course, it can be used as an aid to peer review in these areas (Abramo & D'Angelo, 2011; Abramo et al., 2011; van Raan, 2005b). GS is more comprehensive across all areas but suffers from poor data, especially in terms of multiple versions of the same paper, and also has limitations on data access – no more than 1000 results per query. This particularly affects the calculation of cites per paper (because the number of papers is the divisor) but it does not affect the h-index which only includes the top h papers.

These varied sources do pose the problem that the number of papers and citations may vary significantly and one needs to be aware of this in interpreting any metrics. To illustrate this with a simple example, we have looked up data for one of the authors on WoS and GS. The results are shown in Table 1.

| | **Cites from outputs in WoS using WoS**<br>n, c, h, cpp | **Cites from all sources using GS**<br>n, c, h, cpp |
|---|---|---|
| **Cites to outputs in WoS** | 88, 1684, 21, 19.1 | 87, 4890, 31, 56.2 |
| **Cites to all outputs** | 349, 3796, 30, 10.8 | 316, 13,063, 48, 41.3 |

**Table 1 Comparison of WoS and GS for one of the authors**
n=no. of papers, c=no. of citations, h=h-index (defined below), cpp=cites per paper

The first thing to note is that there are two different ways of accessing citation data in WoS. a) One can do an author search and find all their papers, and then do a citation analysis of those papers. This generates the citations from WoS papers to WoS papers. b) One can do a cited reference search on an author. This generates all the citations from papers in WoS to the author's work whether the cited work is in WoS or not. This therefore generates a much larger number of *cited* publications and a larger number of citations for them. The results are shown in the first column of Table 1. Option a) finds 88 papers in WoS and 1684 citations for them from WoS papers. The corresponding h-index is 21. Option b) finds 349 (!) papers with 3796 citations and an h-index of 30. The 349 papers include many cases of illegitimate duplicates just as does GS. If we repeat the search in GS, we find a total of 316 cited items (cf 349) with 13,063 citations giving an h-index of 48. If we include only the papers that are in WoS we find 87 of the 88, but with 4890 citations and an h-index of 31. So, one could justifiably argue for an h-index ranging from 21 to 48, and a cpp from 10.8 to 56.2.

## 3. METRICS AND THE "LAWS" OF SCIENTOMETRICS

In this section we will consider the main areas of scientometrics analysis – indicators of productivity and indicators of citation impact.

### *3.1. Indicators of productivity*

Some of the very early work, from the 1920s onwards, concerned productivity in terms of the number of papers produced by an author or research unit; the number of papers journals produce on a particular subject; and the number of key words that texts generate. They all point to a similar phenomenon – the Paretian one that a small proportion of producers are responsible for a high proportion of outputs. This also means that the statistical distributions associated with these phenomena are generally highly skewed. It should be said that the original works were quite approximate and actually provided few examples. They have been formalised by later researchers.

Lotka (1926) studied the frequency distribution of numbers of publications per author, concluding that "the number of authors making $n$ contributions is about $1/n^2$ of those making one" from which can be derived de Solla Price's (1963) "square root law" that "half the scientific papers are contributed by the top square root of the total number of scientific authors". So, typically there are 1/4 authors publishing two papers than one; 1/9 publishing three papers and so on. Lotka's Law generates the following distribution:

$$P(X=k) = (6/\pi^2).k^{-2} \quad \text{where } k = 1, 2, \ldots$$

Glänzel and Schubert (1985) showed that a special case of the Waring distribution satisfies the square root law.

Bradford (1934) hypothesised that if one ranks journals in terms of number of articles they publish on a particular subject, then there will be a core that publish the most. If you then group the rest into zones such that each zone has about the same number of articles, then the number of journals in each zone follows this law:

$$N_n = k^n N_0$$

where $k$ = Bradford coefficient, $N_0$ = number in core zone, $N_n$ = journals in the $n^{th}$ zone;

Thus the number of journals needed to publish the same number of articles grows with a power law.

Zipf (1936) studied the frequency of words in a text and postulated that the rank of the frequency of a word and the actual frequency, when multiplied together, are a constant. That is, the number of occurrences is inversely related to the rank of the frequency. In a simple case, the most frequent word will occur twice as often as the second most frequent, and three times as often as the third.

$rf(r) = C$ r is the rank, f(r) is the frequency of that rank, C is a constant

$f(r) = C\ 1/r$

More generally:

$f(r) = \frac{1/r^s}{\sum_1^N \left(\frac{1}{n^s}\right)}$ N is the number of items, s is a parameter

The Zipf distribution has been found to apply in many other contexts such as the size of city by population. All three of these behaviours ultimately rest on the same cumulative advantage mechanisms (SBS) mentioned above and, indeed, under certain conditions all three can be shown to be mathematically equivalent and a consequence of SBS (Egghe, 2005, Chs. 2 and 3).

However, empirical data on the number of publications per year by, for example, a particular author shows that the Lotka distribution by itself is too simplistic as it does not take into account productivity varying over time (including periods of inactivity) or subject. One approach is to model the process as a mixture of distributions (Sichel, 1985). For example, we could assume that the number of papers per year followed a Poisson distribution with parameter λ, but that the parameter itself varied with a particular distribution depending on age, activity, discipline. If we assume that the parameter follows a Gamma distribution, then this mixture results in a negative-binomial which has been found to have a good empirical fit (Mingers & Burrell, 2006). Moreover, this approach (Burrell, 2003) shows that SBS is a consequence of the underlying model.

### *3.2. Indicators of Impact: Citations*

We should begin by noting that the whole idea of the citation being a fundamental indicator of impact, let alone quality, is itself the subject of considerable debate. This concerns: the reasons for citing others' work, Weinstock (1971) lists 15, or not citing it; the meaning or interpretation to be given to citations (Cozzens, 1989; Day, 2014; Leydesdorff, 1998); their place within scientific culture (Wouters, 2014); and the practical problems and biases of citation analysis (Chapman, 1989). This wider context will be discussed later; this section will concentrate on the technical aspects of citation metrics.

The basic unit of analysis is a collection of papers (or more generally research outputs including books reports etc. but as pointed out in Section 2 the main databases only cover journal papers) and the number of citations they have received over a certain period of time. There are three possible situations: a fixed collection observed over a fixed period of time (e.g., computing JIFs); a fixed collection over an extending period of time (e.g., computing JIFs over different time windows); or a collection that is developing over time (e.g., observing the dynamic behaviour of citations over time (Mingers, 2008a)).

*Citation patterns*

If we look at the number of citations per year received by a paper over time it shows a typical birth-death process. Initially there are few citations; then the number increases to a maximum; finally they die away as the content becomes obsolete. Note that the total number of citations can only increase over time but the rate of increase of citations can decrease as obsolescence sets in. There are many variants to this basic pattern, for example "shooting stars" that are highly cited but die quickly, and "sleeping beauties" that are ahead of their time (van Raan, 2004). There are also significantly different patterns of citation behaviour between disciplines that will be discussed in the normalization section. There are several statistical models of this process. Glänzel and Schoepflin (1995) use a linear birth process; Egghe (2000) assumed citations were exponential and deterministic. Perhaps the most usual is to conceptualise the process as basically random from year to year but with some underlying mean ($\lambda$) and use the Poisson distribution. There can then be two extensions – the move from a single paper to a collection of papers with differing mean rates (Burrell, 2001), and the incorporation of obsolescence in the rate of citations (Burrell, 2002; Burrell, 2003).

If we assume a Gamma distribution for the variability of the parameter $\lambda$, then the result is a negative binomial of the form:

$$P(X_t = r) = \binom{r+\nu-1}{\nu-1}\left(\frac{\alpha}{\alpha+t}\right)^{\nu}\left(1-\frac{\alpha}{\alpha+t}\right)^{r}, \quad r = 0, 1, 2, \ldots$$

With mean = $\nu t/\alpha$ variance = $\nu t(t+\alpha)/\alpha^2$ where $\nu$ and $\alpha$ are parameters to be determined empirically.

The negative binomial is a highly skewed distribution which, as we have seen, is generally the case with bibliometric data. Mingers and Burrell (2006) tested the fit on a sample of 600 papers published in 1990 in six MS/OR journals – *Management Science, Operations Research, Omega, EJOR, JORS* and *Decision Sciences* - looking at fourteen years of citations. Histograms are shown in Figure 1 and summary statistics in Table 2. As can be seen, the distributions are highly skewed, and they also have modes (except *ManSci*) at zero, i.e., many papers have never been cited in all that time. The proportion of zero cites varies from 5% in *Management Science* to 22% in *Omega*.

| | *JORS | Omega | EJOR | Dec Sci | Ops Res | Man Sci |
|---|---|---|---|---|---|---|
| Actual mean | 7.3 | 7.2 | 11.3 | 11.1 | 14.6 | 38.6 |
| Actual sd | 17.9 | 15.5 | 19.0 | 14.0 | 28.6 | 42.4 |
| % zero cites | 18 | 22 | 14 | 12 | 10 | 5 |
| Max cites | 176 | 87 | 140 | 66 | 277 | 181 |

**Table 2 Summary statistics for citations in six OR journals 1990-2004, from (Mingers & Burrell, 2006)**

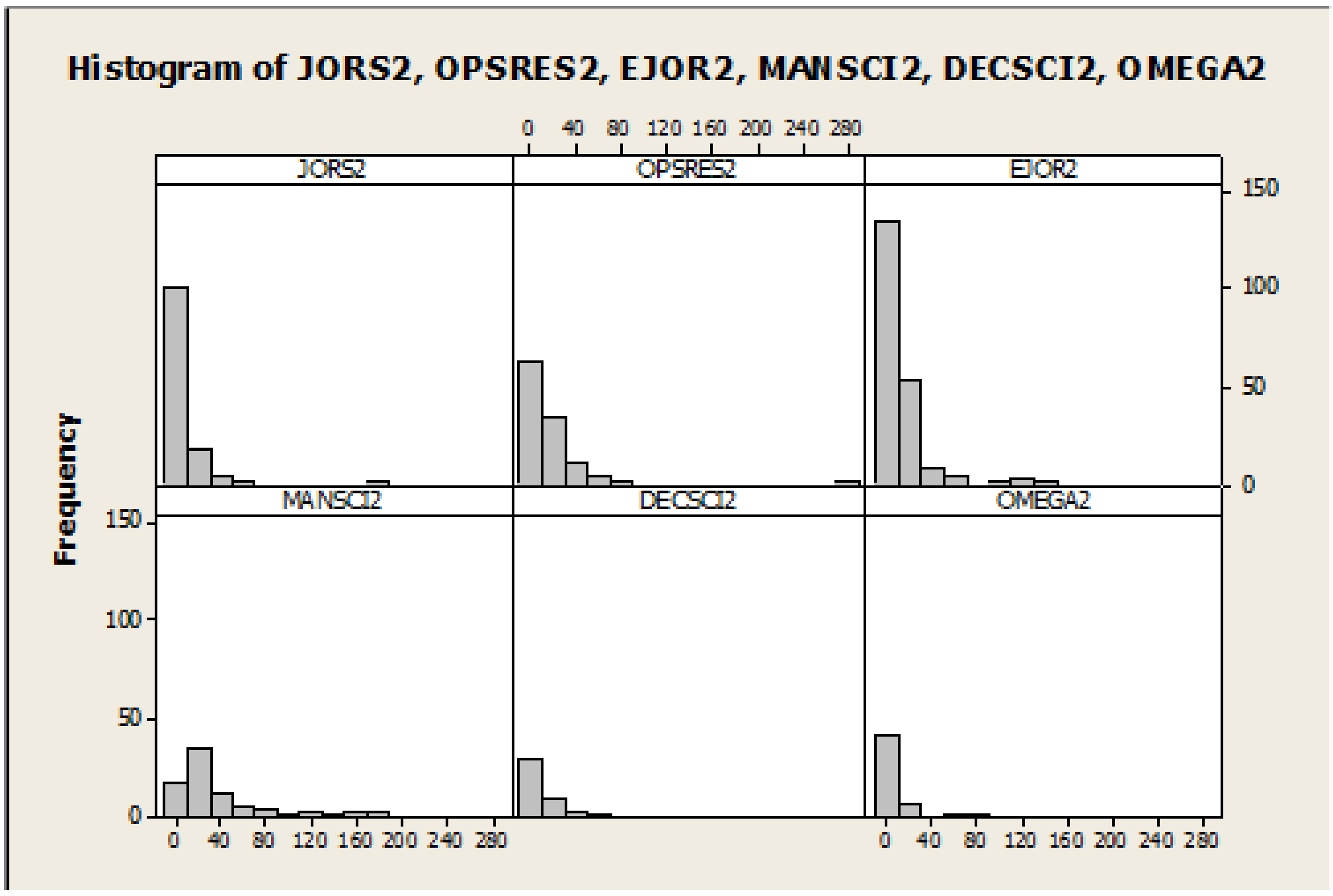

**Figure 1 Histograms for papers published in 1990 in six management science journals, from (Mingers & Burrell, 2006)**

The issue of zero cites is of concern. On the one hand, that a paper has never been cited does not imply that it is of zero quality, especially when it has been through rigorous reviewing processes in a top journal, which is evidence that citations are not synonymous with quality. On the other hand, as Braun (1985) argues, a paper that has never been cited must at the least be disconnected from the field in question. The mean cites per paper (over 15 years) vary considerably between journals from 7.2 to 38.6 showing the major differences between journals (to be covered in a later section), although it is difficult to disentangle whether this is because of the intrinsically better quality of the papers or simply the reputation of the journal. Bornmann et al. (2013a) found that the journal can be considered as a significant co-variate in the prediction of citation impact.

Obsolescence can be incorporated into the model by including a time-based function in the distribution. This would generally be an S-shaped curve that would alter the value of $\lambda$ over time, but there are many possibilities (Meade & Islam, 1998) and the empirical results did not identify any particular one although the gamma and the Weibull distributions provided the best fits. It is also possible to statistically predict how many additional citations will be generated if a particular number have been received so far. The main results are that, at time $t$, the future citations are a linear function of the citations received so far, and the slope of the increment line decreases over the lifetime of the papers. These results applied to collections of papers, but do not seem to apply to the dynamics of individual papers.

In a further study of the same data set, the citation patterns of the individual papers were modelled (Mingers, 2008a). The main conclusions were twofold: i) that individual papers were highly variable and it was almost impossible to predict the final number of citations based on the number in the early years, in fact up to about year ten. This was partly because of sleeping beauty and shooting star effects. ii) The time period for papers to mature was quite long – the maximum citations were not reached until years eight and nine, and many papers were still being strongly cited at the end of 14 years. This is very different from the natural sciences where the pace of citations is very much quicker for most papers (Baumgartner & Leydesdorff, 2014).

If we wish to use citations as a basis for comparative evaluation, whether of researchers, journals or departments, we must consider influences on citations other than pure impact or quality. The first, and most obvious, is simply the number of papers generating a particular total of citations. A journal or department publishing 100 papers per year would expect more citations than one publishing 20. For this reason the main comparative indicator that has been used traditionally is the mean cites per paper (CPP) or raw impact per paper (RIP) during the time period of study. This was the basis of the Leiden (CWTS) "crown indicator" measure for evaluating research units suitably normalised against other factors. (Waltman et al., 2010; Waltman et al., 2011). We should note that this is the opposite of total citations – it pays no attention at all to the number of papers, so a researcher with a CPP of 20 could have one paper, or one hundred papers each with 20 citations.

These other factors include: the general disciplinary area – natural science, social science or humanities; particular fields such as biomedicine (high) or mathematics (low); the type of paper (reviews are high); the degree of generality of the paper (i.e., of interest to a large or small audience); reputational effects such as the journal, the author, or the institution; the language; the region or country (generally the US has the highest number of researchers and therefore citations) as well as the actual content and quality of the paper.

Another interesting issue is whether all citations should be worth the same? There are three distinct factors here – the number of authors of a paper, the number of references in the citing paper, and the quality of the citing journal. In terms of numbers of authors, the sciences generally have many collaborators within an experimental or laboratory setting who all get credited. Comparing this with the situation of a single author who has done all the work themselves, should not the citations coming to that paper be spread among the authors? The extreme example mentioned above concerning the single paper announcing the Higgs Boson actually had a significant effect on the position of several universities in the 2014 Times Higher World University Ranking (Holmes, 2014). The paper, with 2896 "authors" affiliated to 228 institutions, had received 1631 citations within a year. All of the institutions received full credit for this and for some, who only had a relatively small number of papers, it made a huge difference (Aziz & Rozing, 2013; Moed, 2000).

The number of references in the citing paper can be a form of normalisation (fractional counting of citations) (Leydesdorff & Bornmann, 2011a) which will be discussed below. Taking into account the quality of the citing journal gives rise to new indicators that will be discussed in the section on journals.

*The h-index*

We have seen that the total number of citations, as a metric, is strongly affected by the number of papers but does not provide any information on this. At the opposite extreme, the CPP is totally insensitive to productivity. In 2005, a new metric was proposed by Hirsch (2005) that combined in a single, easy to understand, number both impact (citations) and productivity (papers). The h-index has been hugely influential since then, generating an entire literature of its own. Currently his paper has well over 4000 citations in GS. In this section we will only be able to summarise the main advantages and disadvantages, for more detailed reviews see (Alonso et al., 2009; Bornman & Daniel, 2005; Costas & Bordons, 2007; Glänzel, 2006) and for mathematical properties see Glänzel (2006) and Franceschini & Maisano(2010).

The h index is defined as: "a scientist has index $h$ if $h$ of his or her $N_p$ papers have at least $h$ citations each and the other $(N_p - h)$ papers have <= $h$ citations each" (p. 16569).

So $h$ represents the top h papers, all of which have at least $h$ citations. This one number thus combines both number of citations and number of papers. These h papers are generally called the h-core. The h-core is not uniquely defined where more than one paper has h citations. The h-index ignores all the other papers below h, and it also ignores the actual number of citations received above h. The advantages are:

- It combines both productivity and impact in a single measure that is easily understood and very intuitive.
- It is easily calculated just knowing the number of citations either from WoS, Scopus or Google Scholar. Indeed, all three now routinely calculate it.
- It can be applied at different levels – researcher, journal or department.
- It is objective and a good comparator within a discipline where citation rates are similar.
- It is robust to poor data since it ignores the lower down papers where the problems usually occur. This is particularly important if using GS.

However, many limitations have been identified including some that affect all citation based measures (e.g., the problem of different scientific areas, and ensuring correctness of data), and a range of modifications have been suggested (Bornmann et al., 2008).

- The first is that the metric is insensitive to the actual numbers of citations received by the papers in the h-core. Thus two researchers (or journals) with the same h-index could have dramatically different actual numbers of citations. Egghe (2006) has suggested the g-index as a way of compensating for this. "A set of papers has a g-index of *g* if *g* is the highest rank such that the top *g* papers have, together, at least $g^2$ citations" (p. 132). The fundamental idea is that the h-core papers must have at least $h^2$ citations between them although in practice they may have many more. At first sight, the use of the square rather than the cube or any other power seems arbitrary but it is a nice choice since the definition can be re-written so that "the top g papers have an average number of citations at least g", which is much more intuitively appealing. g is at least as large as h.
- The more they have, the larger g will become and so it will to some extent reflect the total number of citations. The disadvantage of this metric is that it is less intuitively obvious than the h-index. Another alternative is the e-index proposed by Zheng (2009). There are several other proposals that measure statistics on the papers in the h-core, for example:
    - The a-index (Jin, 2006; Rousseau, 2006) which is the mean number of citations of the papers in the h-core.
    - The m-index (Bornmann et al., 2008) which is the median of the papers in the h-core since the data is always highly skewed. Currently *Google Scholar Metrics*[13] implements a 5-year h-index and 5-year m-index.

[13] http://scholar.google.co.uk/citations?view_op=top_venues&hl=en

- The r-index (Jin, 2007) which is the square root of the sum of the citations of the h-core papers. This is because the a-index actually penalises better researchers as the number of citations are divided by h, which will be bigger for better scientists. A further development is the ar-index (Jin et al., 2007) which is a variant of the r-index also taking into account the age of the papers.

- The h-index is strictly increasing and strongly related to the time the publications have existed. This biases it against young researchers. It also continues increasing even after a researcher has retired. Data on this is available from Liang(2006)who investigated the actual sequence of h values over time for the top scientists included in Hirsch's sample. A proposed way round this is to consider the h-rate (Burrell, 2007), that is the h-index at time *t* divided by the years since the researcher's first publication. This was also proposed by Hirsch as the *m* parameter in his original paper. Values of 2 or 3 indicate scientists who are both highly productive and well cited.
- The h-index does not discriminate well since it only employs integer values. Given that most researchers may well have h-indexes between 10 and 30, many will share the same value. Guns and Rousseau (2009) have investigated real and rational variants of both *g* and *h*.
- As with all citation-based indicators, they need to be normalised in some way to citation rates of the field. Iglesias and Pecharroman (2007) collected, from WoS, the mean citations per paper in each year from 1995-2005 for 21 different scientific fields The totals ranged from under 2.5 for computer science and mathematics to over 24 for molecular biology. From this data they constructed a table of normalisation factors to be applied to the h-index depending on the field and also the total number of papers published by the researcher. A similar issue concerns the number of authors. The sciences tend to have more authors per paper than the social sciences and humanities and this generates more papers and more citations. Batista et al. (2006) developed the hI-index as the h-index divided by the mean number of authors of the h-core papers. They also claim that this accounts to some extent for the citation differences between disciplines. *Publish or Perish* also corrects for authors by dividing the citations for each paper by the number of authors before calculating the hI,norm-index. This metric has been further normalised to take into account the career length of the author (Harzing et al., 2014).
- The h-index is dependent on or limited by the total number of publications and this is a disadvantage for researchers who are highly cited but for a small number of publications (Costas & Bordons, 2007). For example, Aguillo[14] has compiled a list of the most highly cited researchers in GS according to the h-index (382 with h's of 100 or more). A notable absentee is Thomas Kuhn, one of the most influential researchers of the last 50 years with his concept of a scientific paradigm. His book (Kuhn, 1962) alone has (14/11/14) 74,000 citations which, if the table were ranked in terms of total citations would put him in the top 100. His actual total citations are around 115,000 citations putting him in the top 20. However, his h-index is only 64. This example shows how different metrics can lead to quite extreme results – on the h-index he is nowhere; on total citations, in the top 20; and on cites per paper probably first!

[14] http://www.webometrics.info/en/node/58

There have been many comparisons of the h-index with other indicators. Hirsch himself performed an empirical test of the accuracy of indicators in predicting the future success of researchers and concluded, perhaps unsurprisingly, that the h-index was most accurate (Hirsch, 2007). This was in contrast to other studies such as (Bornmann & Daniel, 2007; Lehmann et al., 2006; van Raan, 2005a). Generally, such comparisons show that the h-index is highly correlated with other bibliometric indicators, but more so with measures of productivity such as number of papers and total number of citations, rather than with citations per paper which is more a measure of pure impact (Alonso et al., 2009; Costas & Bordons, 2007; Todeschini, 2011).

There have been several studies of the use of the h-index in business and management fields such as information systems (Oppenheim, 2007; Truex III et al., 2009), management science (Mingers, 2008b; Mingers et al., 2012), consumer research (Saad, 2006), Marketing (Moussa & Touzani, 2010) and business (Harzing & Van der Wal, 2009).

Overall, the h-index may be somewhat crude in compressing information about a researcher into a single number, and it should always be used for evaluation purposes in combination with other measures or peer judgement but it has clearly become well-established in practice being available in all the citation databases.

Another approach is the use of percentile measures which we will cover in the next section.

## 4. Normalisation Methods

In considering the factors that affect the number of citations that papers receive, there are many to do with the individual paper – content, type of paper, quality, author, or institution (Mingers & Xu, 2010) – but underlying those there are clear disciplinary differences that are hugely significant. As mentioned above, Iglesias and Pecharroman (2007) found that mean citation rates in molecular biology were ten times those in computer science. The problem is not just *between* disciplines but also *within* disciplines such as business and management which encompass different types of research fields. Mingers and Leydesdorff (2014) found that management and strategy papers averaged nearly four times as many citations as public administration. This means that comparisons between researchers, journals or institutions across fields will not be meaningful without some form of normalisation. It is also important to normalise for time period because the number of citations always increases over time (Leydesdorff, Bornmann, Opthof, et al., 2011; Waltman & van Eck, 2013b).

### *4.1. Field Classification Normalisation*

The most well established methodology for evaluating research centres was developed by the *Centre for Science and Technology Studies* (CWTS) at Leiden University and is known as the crown indicator or Leiden Ranking Methodology (LRM) (van Raan, 2005c). Essentially, this method compares the number of citations received by the publications of a research unit over a particular time period with that which would be expected, on a world-wide basis across the appropriate field and for the appropriate publication date. In this way, it normalises the citation rates for the department to rates for its whole field. Typically, top departments may have citation rates that are three or four times the field average. Leiden also produces a ranking of world universities based on bibliometric methods that will be discussed elsewhere (Waltman et al., 2012).

This is the traditional "crown indicator", but this approach to normalisation has been criticised (Leydesdorff & Opthof, 2011; Lundberg, 2007; Opthof & Leydesdorff, 2010) and an alternative has been used in several cases (Cambell et al., 2008; Rehn & Kronman, 2008; Van Veller et al., 2009). This has generated considerable debate in the literature (Bornmann, 2010; Bornmann & Mutz, 2011; Moed, 2010a; van Raan et al., 2011; Waltman et al., 2010; Waltman et al., 2011). The main criticism concerns the order of calculation in the indicator but the use of a mean when citation distributions are highly skewed is also a concern. It is argued that, mathematically, it is wrong to sum the actual and expected numbers of citations separately and then divide them. Rather, the division should be performed first, for each paper, and then these ratios should be averaged. In the latter case you get a proper statistic rather than merely a quotient. It might be thought that this is purely a technical issue, but it it can affect the results significantly. In particular, the older CWTS method tends to weight more highly publications from fields with high citation numbers whereas the new one weights them equally. Also, the older method is not consistent in its ranking of institutions when both improve equally in terms of publications and citations. Eventually this was accepted by CWTS, and Waltman et al. (2010; 2011) (from CTWS) have produced both theoretical and empirical comparisons of the two methods and concluded that the newer one is theoretically preferably but does not make much difference in practice. The new method is called the "mean normalised citation score" (MNCS). Gingras et al. (2011) commented that the "alternative" method was not alternative but in fact the correct way to normalise, and had been in use elsewhere for fifteen years.

### *4.2. Source Normalisation*

The normalisation method just discussed normalised citations against other citations, but an alternative approach was suggested, initially by Zitt and Small (2008) in their "audience factor", which considers the *sources* of citations, that is the reference lists of citing papers, rather than citations themselves. This general approach is gaining popularity and is also known as the "citing-side approach" (Zitt, 2011), source normalisation (Moed, 2010c) (SNIP), fractional counting of citations (Leydesdorff & Opthof, 2010) and a priori normalisation (Glänzel et al., 2011).

The essential difference in this approach is that the reference set of journals is not defined in advance, according to WoS or Scopus categories, but rather is defined at the time specifically for the collection of papers being evaluated (whether that is papers from a journal, department, or individual). It consists of all the papers, in the given time window, that cite papers in the target set. Each collection of papers will, therefore, have its own unique reference set and it will be the lists of references from those papers that will be used for normalisation. This approach has obvious advantages – it avoids the use of WoS categories which are ad hoc and outdated (Leydesdorff & Bornmann, 2014; Mingers & Leydesdorff, 2014) and it allows for journals that are interdisciplinary and that would therefore be referenced by journals from a range of fields.

Having determined the reference set of papers, the methods then differ in how they employ the number of references in calculating a metric. The audience factor (Zitt, 2011; Zitt & Small, 2008) works at the level of a citing journal. It calculates a weight for citations from that journal based on the ratio between the average number of active references[15] in all journals to the average number of references in the citing journal. This ratio will be larger for journals that have few references compared to the average because they are in less dense citation fields. Citations to the target (cited) papers are then weighted using the calculated weights which should equalise for the citation density of the citing journals.

Fractional counting of citations (Leydesdorff & Bornmann, 2011a; Leydesdorff & Opthof, 2010; Leydesdorff, Radicchi, et al., 2013; Small & Sweeney, 1985; Zitt & Bassecoulard, 1994) begins at the level of an individual citation and the paper which produced it. Instead of counting each citation as one, it counts it as a fraction of the number of references in the citing paper. This if a citation comes from a paper with *m* references, the citation will have a value of $1/m$. It is then legitimate to add all these fractionated citations to give the total citation value for the cited paper. An advantage of this approach is that statistical significance tests can be performed on the results. One issue is whether all references are included (which Leydesdorff et al. do) or whether only the active references should be counted. The third method is essentially that which underlies the SNIP indicator for journals (Moed, 2010b) which will be discussed in Section 5. In contrast to fractional counting, it forms a ratio of the mean number of citations to the journal to the mean number of references in the citing journals. A later version of SNIP (Waltman et al., 2013) used the harmonic mean to calculate the average number of references and in this form it is essentially the same as fractional counting except for an additional factor to take account of papers with no active citations.

---

[15] An "active" reference is one that is to a paper included in the database (e.g., WoS) within the time window. Other references are then ignored as "non-source references".

Some empirical reviews of these approaches have been carried out. Waltman and van Eck (2013a; Waltman & van Eck, 2013b) compared the three source-normalising methods with the new CWTS crown indicator (MNCS) and concluded that the source normalisation methods were preferable to the field classification approach, and that of them, the audience factor and revised SNIP were best. This was especially noticeable for interdisciplinary journals. The fractional counting method did not fully eliminate disciplinary differences (Radicchi & Castellano, 2012) and also did not account for citation age.

### *4.3. Percentile-Based Approaches*

We have already mentioned that there is a general statistical problem with metrics that are based on the mean number of citations, which is that citations distributions are always highly skewed (Seglen, 1992) and this invalidates the mean as a measure of central tendency; the median is better. There is also the problem of ratios of means discussed above. A non-parametric alternative based on percentiles (an extension of the median) has been suggested for research groups (Bornmann & Mutz, 2011), individual scientists (Leydesdorff, Bornmann, Mutz, et al., 2011) and journals (Leydesdorff & Bornmann, 2011b). This is also used by the US National Science board in their *Science and Engineering Indicators*.

The method works as follows:

1. For each paper to be evaluated, a reference set of papers published in the same year, of the same type and belonging to the same WoS category is determined.
2. These are rank ordered and split into percentile rank (PR) classes, for example the top 1% (99$^{th}$ percentile), 5%, 10%, 25%, 50% and below 50%. For each PR, the minimum number of citations necessary to get into the class is noted[16]. Based on its citations, the paper is then assigned to one of the classes. This particular classification is known as 6PR.
3. The procedure is repeated for all the target papers and the results are then summated, giving the overall percentage of papers in each of the PR classes. The resulting distributions can be statistically tested against both the field reference values and against other competitor journals or departments[17].

The particular categories used above are only one possible set (Bornmann, Leydesdorff & Mutz, 2013) – others in use are [10%, 90%] and [0.01%, 0.1%, 1%, 10%, 20%, 50%] (used in *ISI Essential Science Indicators*) and the full 100 percentiles (100PR) (Bornmann, Leydesdorff, et al., 2013b; Leydesdorff, Bornmann, Mutz, et al., 2011). This approach provides a lot of information about the proportions of papers at different levels, but it is still useful to be able to summarise performance in a single value. The suggested method is to calculate a mean of the ranks weighted by the proportion of papers in each. The minimum is 1, if all papers are in the lowest rank; the maximum is 6 if they are all in the top percentile. The field average will be 1.91 - (.01, .04, 05, .15, .25, .50) x (6,5,4,3,2,1) - so a value above that is better than the field average. A variation of this metric has been developed as an alternative to the journal impact factor (JIF) called *I3* (Leydesdorff, 2012; Leydesdorff & Bornmann, 2011b). Instead of multiplying the percentile ranks by the proportion of papers in each class, they are multiplied by the actual numbers of papers in each class thus giving a measure that combines productivity with citation impact. In the original, the 100PR classification was used but other ones are equally possible.

[16] There are several technical problems to be dealt with in operationalising these classes (Bornmann, Leydesdorff & Mutz, 2013; Bornmann, Leydesdorff, et al., 2013b).
[17] Using Dunn's test or the Mann-Whitney U test (Leydesdorff, Bornmann, Mutz, et al., 2011).

The main drawback of this method is that it relies on the field definitions in WoS or another database which are unreliable, especially for interdisciplinary journals. It might be possible to combine it with some form of source normalisation (Colliander, 2014).

## 5. Indicators of Journal Quality: The Impact Factor and Other Metrics

So far, we have considered the impact of individual papers or researchers, but of equal importance is the impact of journals in terms of library's decisions about which journals to take (less important in the age of e-journals), authors' decisions about where to submit their papers, and in subsequent judgements of the quality of the paper. Indeed journal ranking lists such as the UK *Association of Business Schools*' (ABS) has a huge effect on research behaviour (Mingers & Willmott, 2013). Until recently, the journal impact factor (JIF) has been the pre-eminent measure. This was originally created by Garfield and Sher (1963) as a simple way of choosing journals for their SCI but, once it was routinely produced in WoS (who have copyright to producing it), it became a standard. Garfield recognised its limitations and also recommended a metric called the "cited half-life" which is a measure of how long citations last for. Specifically, it is the median age of papers cited in a particular year, so a journal that has a cited half-life of five years means that 50% of the citations are to papers published in the last five years.

JIF is simply the mean citations per paper for a journal over a two year period. For example, the 2014 JIF is the number of citations in 2014 to papers published in a journal in 2012 and 2013, divided by the number of such papers. WoS also published a 5-year JIF because in many disciplines two years is too short a time period. It is generally agreed that the JIF has few benefits for evaluating research, but many deficiencies (Brumback, 2008; Cameron, 2005; Seglen, 1997; Vanclay, 2012). Even Garfield (1998) has warned about its over-use[18].

- JIF depends heavily on the research field. As we have already seen, there are large differences in the publishing and citing habits of different disciplines and this is reflected in huge differences in JIF values. Looking at the WoS journal citation reports 2013, in the area of cell biology the top journal has a JIF of 36.5 and the 20th one of 9.8. *Nature* has a JIF of 42.4. In contrast, in the field of management, the top journal (*Academy of Management Review*) is 7.8 and the 20th is only 2.9. Many journals have JIFs of less than 1. Thus, it is not appropriate to compare JIFs across fields (even within business and management) without some form of additional normalisation
- The two-year window. This is a very short time period for many disciplines, especially given the lead time between submitting a paper and having it published which may itself be two years. In management, many journals have a cited half-life of over 10 years while in cell biology it is typically less than 6. The 5-year JIF is better in this respect (Campanario, 2011).
- There is a lack of transparency in the way the JIF is calculated and this casts doubt on the results. Brumback (2008) studied medical journals and could not reproduce the appropriate figures. It is highly dependent on which types of papers are included in the denominator. In 2007, the editors of three prestigious medical journals published a paper questioning the data (Rossner et al., 2007). Pislyakov (2009) has also found differences between JIFs calculated in WoS and Scopus for economics resulting from different journal coverage.
- It is possible for journals to deliberately distort the results by, for example, publishing many review articles which are more highly cited; publishing short reports or book reviews that get cited but are not included in the count of papers; or pressuring authors to gratuitously reference excessive papers from the journal (Wilhite & Fong, 2012). The *Journal of the American College of Cardiology*, for example, publishes each year an overview of highlights in its previous year so that the IF of this journal is boosted (DeMaria et al., 2008).
- If used for assessing individual researchers or papers the JIF is unrepresentative (Oswald, 2007). As Figure 1 shows, the distribution of citations within a journal is highly skewed and so the mean value will be distorted by a few highly cited papers, and not represent the significant number that may never be cited at all.

In response to criticisms of the JIF, several more sophisticated metrics have been developed, although the price for sophistication is complexity of calculation and a lack of intuitiveness in what it means.

---

[18] There was a special issue of Scientometrics (92, 2, 2012) devoted to it and also a compilation of 40 papers published in Scientometrics (Braun, 2007)

The first metrics we will consider take into account not just the quantity of citations but also their quality in terms of the prestige of the citing journal. They are based on iterative algorithms over a network, like Googles's PageRank, that initially assign all journals an equal amount of prestige and then iterate the solution based on the number of citations (the links) between the journals (nodes) until a steady state is reached. The first such was developed by Pinsky and Narin (1976) but that had calculation problems. Since then, Page et al. (1999) and Ma (2008) have an algorithm based directly on PageRank but adapted to citations; Bergstrom (2007) has developed the Eigenfactor which is implemented in WoS; and Gonzalez-Pereira et al. (2010) have developed SCImago Journal Rank (SJR) which is implemented in Scopus.

The Eigenfactor is based on the notion of a researcher taking a random walk following citations from one paper to the next, measuring the relative frequency of occurrence of each journal as a measure of prestige. It explicitly excludes journal self-citations unlike most other metrics. Its values tend to be very small, for example the largest in the management field is Management Science with a value of 0.03 while the 20$^{th}$ is 0.008 which is not very meaningful. The Eigenfactor measures the total number of citations and so is affected by the total number of papers published by a journal. A related metric is the article influence score (AI) which is the Eigenfactor divided by the proportion of papers in the database belonging to a particular journal over five years. It can therefore be equated to a 5-yr JIF. A value of 1.0 shows that the journal has average influence; values greater than 1.0 show greater influence. We can see that in cell biology the largest AI is 22.1 compared with 6.6 in management. Fersht (2009) and Davies (2008) argue empirically, that the Eigenfactor gives essentially the same information as total citations as it is size-dependent, but West et al. (2010) dispute this. It is certainly the case that the rankings of journals with the Eigenfactor, which is not normalised for the number of publications, are significantly different to those based on total citations, JIF or AI, which are all quite similar (Leydesdorff, 2009).

| Metric | Description | Advantages | Disadvantages | Maximum values for:<br>a) cell biology<br>b) management | No of papers | field | prestige |
|---|---|---|---|---|---|---|---|
| Impact factor (JIF) and cited half-life (WoS) | Mean citations per paper over a 2 or 5 year window. Normalised to number of papers. Counts citations equally | Well-known, easy to calculate and understand. | Not normalised to discipline; short time span; concerns about data and manipulation | From WoS<br>a) 36.5<br>b) 7.8 | Y | N | N |
| Eigenfactor and article influence score (AI) (WoS) | Based on PageRank, measures citations in terms of the prestige of citing journal. Not normalised to discipline or number of papers. Correlated with total citations. Ignores self-citations. AI is normalised to number of papers, so is like a JIF<br>5-yr window | Very small values, difficult to interpret, not normalised | The AI is normalised to number of papers. A value of 1.0 shows average influence across all journals | From WoS<br>Eigenfactor:<br>a)0.599<br>b)0.03<br><br>AI:<br>a) 22.2<br>b) 6.56 | N<br><br>Y | N<br><br>N | Y<br><br>Y |
| SJR and SJR2 (Scopus) | Based on citation prestige but also includes a size normalisation factor.<br>SJR2 also allows for the closeness of the citing journal.<br>3-year window | Complex calculations and not easy to interpret. Not field normalised | Normalised number of papers but not to field so comparable to JIF. Most sophisticated indicator | | Y | N | Y |
| h-index (Scimago website and Google Metrics) | The h papers of a journal that have at least h citations. Can have any window – Google metrics uses 5-year | Easy to calculate and understand. Robust to poor data | Not normalised to number of papers or field.<br>Not pure impact but includes volums | From Google Metrics<br>h5:<br>a) 223<br>b) 72<br><br>h5 median:<br>a) 343<br>b)122 | N | N | N |
| SNIP<br><br>Revised SNIP (Scopus) | Citations per paper normalised to the relative database citation potential, that is the mean number of references in the papers that cite the journal | Normalises both to number of papers and field. | Does not consider citation prestige. Complex and difficult to check.<br>Revised version is sensitive to variability of number of references | | Y | Y | N |
| I3 | Combines the distribution of citation percentiles with respect to a reference set with the number of papers in each percentile class | Normalises across fields. Does not use the mean but is based on percentiles which is better for skewed data | Needs reference sets based on pre-defined categories such as WoS | Not known | N | Y | N |

**Table 2 Characteristics of Metrics for Measuring Journal Impact**

The SJR works in a similar way to the Eigenfactor but includes within it a size normalisation factor and so is more akin to the article influence score. Each journal is a node and each directed connection is a normalised value of the number of citations from one journal to another over a three year window. It is normalised by the total number of citations in the citing journal for the year in question. It works in two phases:

1. An un-normalised value of journal prestige is calculated iteratively until a steady state is reached. The value of prestige actually includes three components: A fixed amount for being included in the database (Scopus); an amount dependent on the number of papers the journal produces; a citation amount dependent on the number of citations received, and the prestige of the sources. However, there are a number of arbitrary weighting constants in the calculation.
2. The value from 1., which is size-dependent, is then normalised by the number of published articles and adjusted to give an "easy-to-use" value.

Gonzales-Pereira et al. (2010) carried out extensive empirical comparisons with a 3-yr JIF (on Scopus data). The main conclusions were that the two were highly correlated, but the SJR showed that some journals with high JIFs and lower SJRs were indeed gaining citations from less prestigious sources. This was seen most clearly in the computer science field where the top ten journals, based on the two metrics, were entirely different except for the number one, which was clearly a massive outlier (*Briefings in Bioinformatics*). Values for the JIF are significantly higher than for SJR. Falagas et al. (Falagas et al., 2008) also compared the SJR favourably with the JIF.

There are several limitations of these $2^{nd}$ generation measures: the values for "prestige" are difficult to interpret as they are not a mean citation value but only make sense in comparison with others; they are still not normalised for subject areas (Lancho-Barrantes et al., 2010); and the subject areas themselves are open to disagreement (Mingers & Leydesdorff, 2014).

A further development of the SJR indicator has been produced (Guerrero-Bote & Moya-Anegón, 2012) with the refinement that, in weighting the citations according to the prestige of the citing journal, the relatedness of the two journals is also taken into account. An extra term is added based on the cosine of the angle between the co-citation vectors of the journals so that the citations from a journal in a highly related area count for more. It is claimed that this also goes some way towards reducing the disparity of scores between subjects. However, it also makes the indicator even more complex, hard to compute, and less understandable.

The h-index can also be used to measure the impact of journals as it can be applied to any collection of cited papers (Braun et al., 2006; Schubert & Glänzel, 2007; Xu et al., 2015).Studies have been carried out in several disciplines: marketing (Moussa & Touzani, 2010), economics (Harzing & Van der Wal, 2009), information systems (Cuellar et al., 2008) and business (Mingers et al., 2012). The advantages and disadvantages of the h-index for journals are the same as the h-index generally, but it is particularly the case that it is not normalised for different disciplines, and it is also strongly affected by the number of papers published. So a journal that publishes a small number of highly cited papers will be disadvantaged in comparison with one publishing many papers, even if not so highly cited. Google Metrics (part of Google Scholar) uses a 5-year h-index and also shows the median number of citations for those papers in the h core to allow for differences between journals with the same h-index. It has been critiqued by Delgado-López-Cózar and Cabezas-Clavijo (2012).

Another recently developed metric that is implemented in Scopus but not WoS is SNIP – source normalised impact per paper (Moed, 2010b). This normalises for different fields based on the citing-side form of normalisation discussed above, that is, rather than normalising with respected to the *citations* that a journal receives, it normalises with respect to the number of *references* in the citing journals. The method proceeds in three stages:

1. First the raw impact per paper (RIP) is calculated for the journal. This is essentially a three year JIF – the total citations from year *n* to papers in the preceding three years is divided by the number of citable papers.
2. Then the database citation potential for the journal (DCP) is calculated. This is done by finding all the papers in year *n* that cite papers the journal over the preceding ten years, and then calculating the arithmetic mean of the number of references (to papers in the database – Scopus) in these papers.
3. The DCP is then relativized (RDCP). The DCP is calculated for *all* journals in the database and the median value is found. Then $RDCP_j = DCP_j$/Median DCP. Thus a field that has many references will have an RDCP above 1.
4. Finally, $SNIP_j = RIP_j / RDCP_j$

The result is that journals in fields that have a high citation potential will have their RIP reduced, and vice versa for fields with low citation potential. This is an innovative measure both because it normalises for both number of publications and field, and because the set of reference journals are specific to each journal rather than being defined beforehand somewhat arbitrarily. Moed presents empirical evidence from the sciences that the subject normalisation does work even at the level of pairs of journals in the same field. Also, because it only uses references to papers within the database, it corrects for coverage differences – a journal with low database coverage will have a lower DCP and thus a higher value of SNIP.

A modified version of SNIP has recently been introduced (Waltman et al., 2013) to overcome certain technical limitations, and also in response to criticism from Leydesdorff and Opthof (2010; Moed, 2010c) who favour a fractional citation approach. The modified version involves two main changes: i) the mean number of references (DCP), but not the RIP, is now calculated using the harmonic mean rather than the arithmetic mean. ii) The relativisation of the DCP to the overall median DCP is now omitted entirely, now SNIP = RIP/DCP.

Mingers (2014) has pointed out two problems with the revised SNIP. First, because the value is no longer relativized it does not bear any particular relation to either the RIP for a journal, or the average number of citations/references in the database which makes it harder to interpret. Second, the harmonic mean, unlike the arithmetic, is sensitive to the variability of values. The less even the numbers of references, the lower will be the harmonic mean and this can make a significant difference to the value of SNIP which seems inappropriate. There is also a more general problem with these sophisticated metrics that work across a whole database, and that is that the results cannot be easily replicated as most researchers do not have sufficient access to the databases (Leydesdorff, 2013).

Two other alternatives to the JIF have been suggested (Leydesdorff, 2012) – fractional counting of citations, which is similar in principle to SNIP, and the use of non-parametric statistics such as percentiles which avoids using means which are inappropriate with highly skewed data. A specific metric, based on percentiles, called I3 has been proposed by Leydesdorff (2011b) which combines relative citation impact with productivity in terms of the numbers of papers but is normalised through the use of percentiles (see Section 4.3 for more explanation).

## 6. Visualizing and mapping science

In addition to its use as an instrument for the evaluation of impact, citations can also be considered as an operationalization of a core process in scholarly communication, namely, referencing. Citations refer to texts other than the one that contains the cited references, and thus induce a dynamic vision of the sciences developing as networks of relations (Price, 1965). The development of co-citation analysis (Marshakova, 1973; Small, 1973) and co-word analysis (Callon et al., 1983) were achievements of the 1970s and 1980s. Aggregated journal-journal citations are available on a yearly basis in the Journal Citation Reports of the Science Citation Index since 1975. During the mid-80s several research teams began to use this data for visualization purposes using multidimensional scaling and other such techniques (Doreian & Fararo, 1985; Leydesdorff, 1986; Tijssen et al., 1987). The advent of graphical user-interfaces in Windows during the second half of the 1990s stimulated the further development of network analysis and visualization programs such as Pajek (de Nooy et al., 2011) that enable users to visualize large networks. Using large computer facilities, Boyack *et al.*. (2005) first mapped "the backbone" of all the sciences (De Moya-Anegón et al., 2007).

Bollen *et al.*. (2009) developed maps based on clickstream data; Rosvall & Bergstrom (2010) proposed to use alluvial maps for showing the dynamics of science. Rafols *et al.*. (2010) first proposed to use these "global" maps as backgrounds for overlays that inform the user about the position of specific sets of documents, analogously to overlaying institutional address information on geographical maps like Google Maps. More recently, these techniques have been further refined, using both journal (Leydesdorff, Rafols, et al., 2013) and patent data (Kay et al., 2014)).

Nowadays, scientometric tools for the visualization are increasingly available on the internet. Some of them enable the user directly to map input downloaded from Web of Science or Scopus. VOSviewer[19] (Van Eck & Waltman, 2010) can generate, for example, co-word and co-citation maps from this data.

### *6.1. Visualisation techniques*

The systems view of multidimensional scaling (MDS) is deterministic, whereas the graph-analytic approach can also begin with a random or arbitrary choice of a starting point. Using MDS, the network is first conceptualized as a multi-dimensional space that is then reduced stepwise to lower dimensionality. At each step, the stress increases. Kruskall's stress function is formulated as follows:

$$S = \sqrt{\frac{\sum_{i \neq j} (\|x_i - x_j\| - d_{ij})^2}{\sum_{i \neq j} d_{ij}^2}} \qquad (1)$$

In this formula $\|x_i - x_j\|$ is equal to the distance on the map, while the distance measure $d_{ij}$ can be, for example, the Euclidean distance in the data under study. One can use MDS to illustrate factor-analytic results in tables, but in this case the Pearson correlation is used as the similarity criterion.

Spring-embedded or force-based algorithms can be considered as a generalization of MDS, but were inspired by developments in graph theory during the 1980s. Kamada and Kawai (1989) were the first to reformulate the problem of achieving target distances in a network in terms of energy optimization. They formulated the ensuing stress in the graphical representation as follows:

[19] http://www.vosviewer.com

$$S = \sum_{i \neq j} s_{ij} \quad \text{with} \quad s_{ij} = \frac{1}{d_{ij}^{\;2}} (\|x_i - x_j\| - d_{ij})^2 \qquad (2)$$

Equation 2 differs from Equation 1 by taking the square root in Equation 1, and because of the weighing of *each* term with $1/d_{ij}^2$ in the numerator of Equation 2. This weight is crucial for the quality of the layout, but defies normalization with $\sum d_{ij}^2$ in the denominator of Equation 1; hence the difference between the two stress values. Note that 1 is a ratio of sums while 2 is a sum of ratios (see discussion above).

The ensuing difference at the conceptual level is that spring-embedding is a graph-theoretical concept developed for the topology of a network. The weighting is achieved for each individual link. MDS operates on the multivariate space as a system, and hence refers to a different topology. In the multivariate space, two points can be close to each other without entertaining a relationship. For example, they can be close or distanced in terms of the correlation between their *patterns* of relationships.

In the network topology, Euclidean distances and geodesics (shortest distances) are conceptually more meaningful than correlation-based measures. In the vector space, correlation analysis (factor analysis, etc.) is appropriate for analysing the main dimensions of a system. The cosines of the angles among the vectors, for example, build on the notion of a multi-dimensional space. In bibliometrics, Ahlgren et al. (2003) have argued convincingly in favour of the cosine as a non-parametric similarity measure because of the skewedness of the citation distributions and the abundant zeros in citation matrices. Technically, one can also input a cosine-normalized matrix into a spring-embedded algorithm. The value of (1 – *cosine*) can then be considered as a distance in the vector space (Leydesdorff & Rafols, 2011).

Newman & Girvan (2004) developed an algorithm in graph theory that searches for (latent) community structures in networks of observable relations. An objective function for the decomposition is recursively minimized and thus a "modularity" *Q* can be measured (and normalized between zero and one). Blondel et al. (2008) improved community-finding for large networks; this routine is implemented in Pajek and Gephi, whereas Newman & Girvan's original routine can be found in the Sci2 Toolset for "the science of science".[20] VOSviewer provides its own algorithms for the mapping and the decomposition .

[20] https://sci2.cns.iu.edu/user/index.php

## *6.2. Local and global maps*

To illustrate some of these possibilities, we analysed the 505 documents published in the *European Journal of Operational Research* in 2013[21]. Among the 1,555 non-trivial words in the titles of these documents, 58 words occur more than ten times and form a large component. A semantic map of these terms is shown in Figure 2.

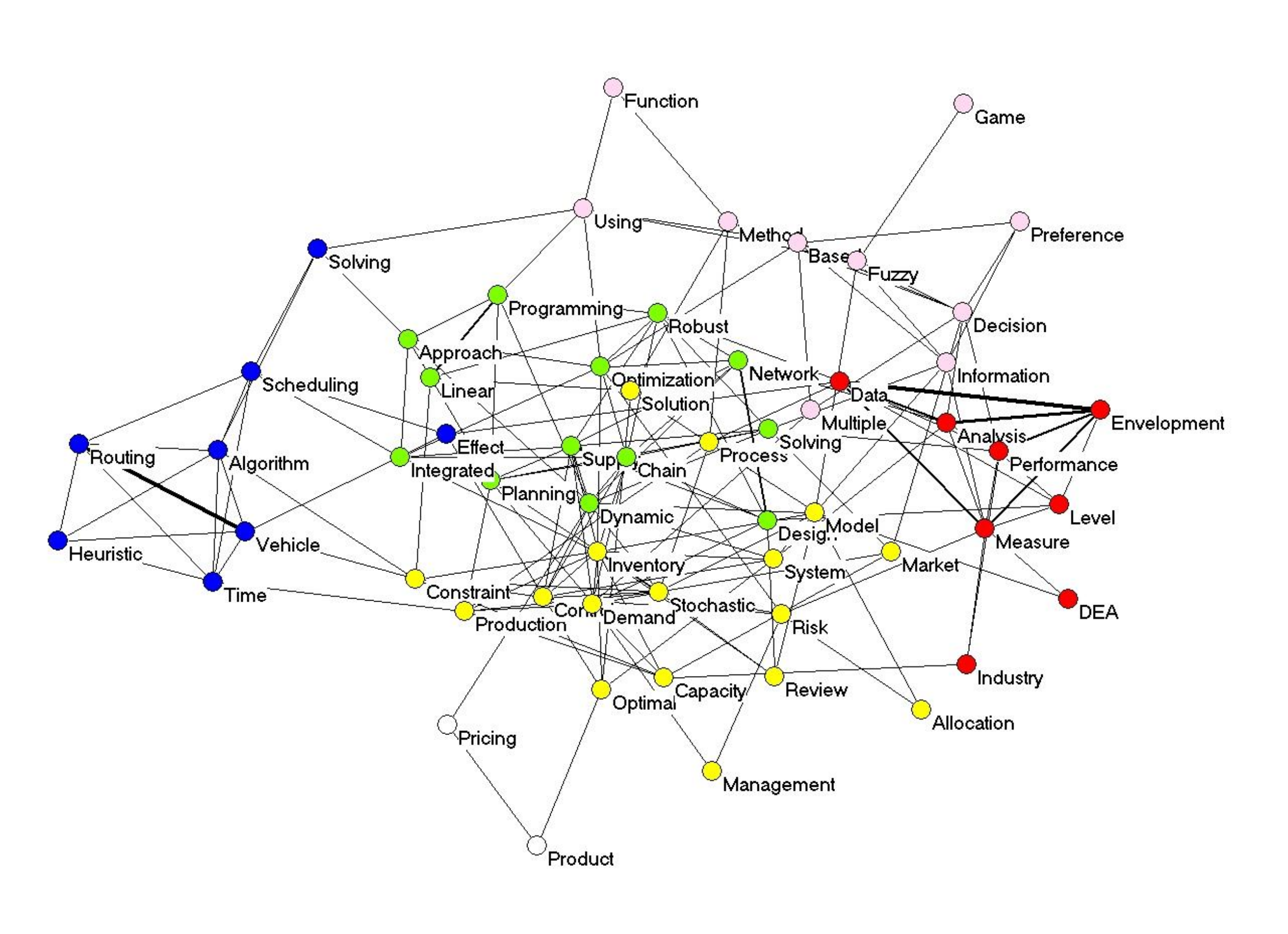

**Figure 2: Cosine-normalized map of the 58 title words which occur ten or more times in the 505 documents published in EJOR during 2013.** (cosine > 0.1; modularity Q = 0.548 using Blondel et al.., (2008); Kamada & Kawai (1989) used for the layout; see http://www.leydesdorff.net/software/ti.)

In Figure 2 we can see some sensible groupings – for example transportation/scheduling, optimization/programming, decision analysis, performance measurement and a fifth around management/application areas.

Figure 3 shows the 613 journals that are most highly cited in the same 505 *EJOR* papers (12,172 citations between them) but overlaid on to a global map of science (Leydesdorff, Rafols, et al., 2013). These cited sources can, for example, be considered as an operationalization of the knowledge base on which these articles draw. It can be seen that, apart from the main area around OR and management, there is significant citation to the environmental sciences, chemical engineering, and biomedicine. Rao-Stirling diversity — a measure for the interdisciplinarity of this knowledge base (Rao, 1982) — however, is low (0.1187). In other words, citation within the specialty prevails.

[21] Using a routine available at http://www.leydesdorff.net/software/ti

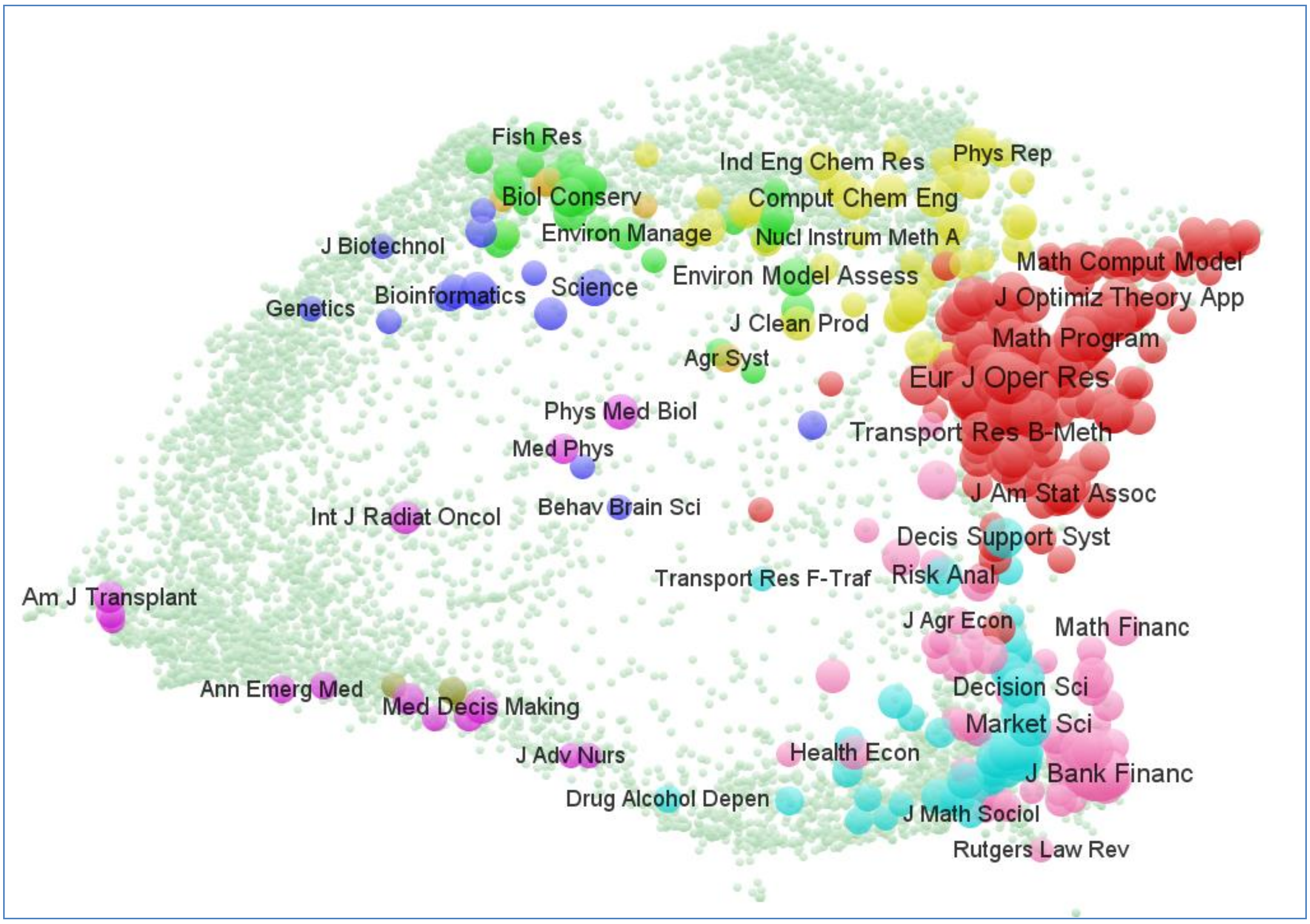

**Figure 3: 613 journals cited in 505 documents published in EJOR during 2013, overlaid on the global map of science in terms of journal-journal citation relations.** (Rao-Stirling diversity is 0.1187; Leydesdorff et al.. (in press); see at http://www.leydesdorff.net/journals12 ).

Figure 4 shows a local map of the field of OR based on the 29 journals most highly cited in papers published in *Operations Research* in 2013. In this map three groupings have emerged – the central area of OR, including transportation; the lower left of particularly mathematical journals; and the upper region of economics and finance journals which includes *Management Science*.

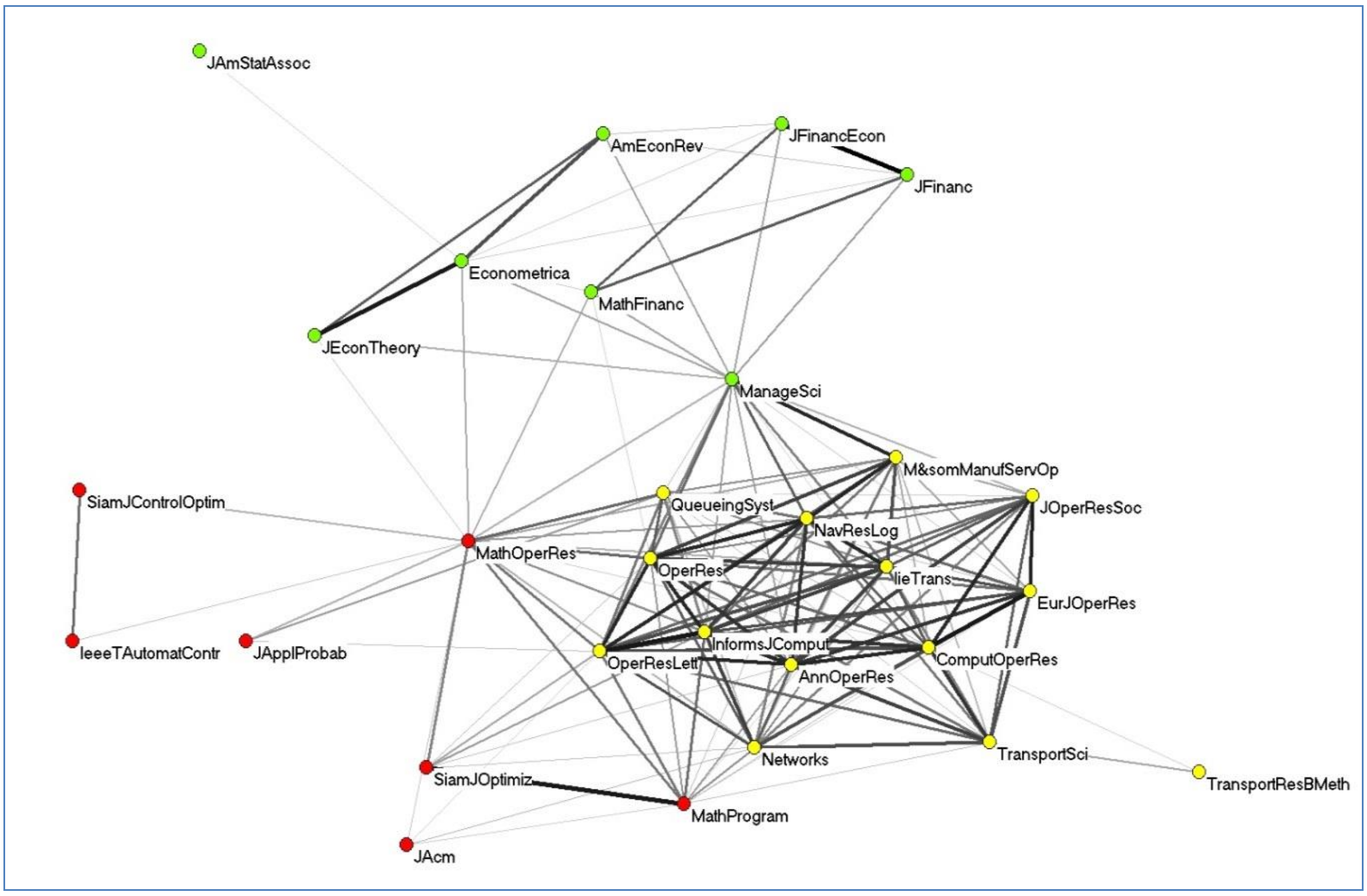

**Figure 4: local map of the 29 journals cited in articles of Operations Research in 2013** (1% level; cosine > 0.2; Kamada & Kawai, 1989; Blondel et al.., 2008; Q = 0.213).

In summary, the visualizations enable us to represent the current state of the field (Figure 2), its knowledge base (Figure 3), and its relevant environments (Figure 4). Second-order visualization programs available at the internet such as *VOSviewer* and *CitNetExplorer*[22] enable the user to automatically generate several of these visualizations from data downloaded from WoS or Scopus. One can also envisage making movies from this data. These networks evolve over time and the diagrams can be animated – see, for example: http://www.leydesdorff.net/journals/nanotech/ or other ones at http://www.leydesdorff/visone for an overview and instruction.

## 7. Evaluation and Policy

As we said in the introduction, scientometrics has come to prominence because of its use in the evaluation and management of research performance, whether at the level of the researcher, research group, institution or journal (Bornmann & Leydesdorff, 2014). Many countries, especially the UK, Australia, New Zealand and Italy, carry out regular reviews of university performance affecting both the distribution of research funds and the generation of league tables. On a macro scale, world university league tables have proliferated (e.g., QS[23], Times Higher[24] and Shanghai[25]) including one from Leiden[26] based purely on bibliometrics (Waltman et al., 2012) while on a micro scale personal employment and promotion is shaped by journal and citation data. Much of this is based on the citation metrics that we have discussed above.

[22] http://www.citnetexplorer.nl/Home
[23] http://www.topuniversities.com/
[24] http://www.timeshighereducation.co.uk/world-university-rankings/

The traditional method of research evaluation was peer review (Abramo et al., 2011; Bornmann, 2011; Moed, 2007). However, this has many drawbacks – it is very time consuming and costly (Abramo & D'Angelo, 2011), subject to many biases and distortions (Horrobin, 1990; Moxham & Anderson, 1992), generally quite opaque (panel members in the 2008 UK RAE were ordered to destroy all notes for fear of litigation) (Reale et al., 2007) and limited in the extent to which it actually provides wide-ranging and detailed information (Butler & McAllister, 2009). The UK did investigate using bibliometrics in 2008 (HEFCE, 2009), used them to a limited extent in 2014, and are expected to employ them more fully in 2020. In contrast, bibliometrics has the potential to provide a cheaper, more objective and more informative mode of analysis, although there is general agreement that bibliometrics should only be used in combination with some form of transparent peer review Moed (2007; van Raan, 2005b). Abramo and D'Angelo (2011) compared informed peer review (including the UK RAE) with bibliometrics on the natural and formal sciences in Italy and concluded that bibliometrics were clearly superior across a range of criteria – accuracy, robustness, validity, functionality, time and cost. They recognized that there were problems in the social sciences and humanities where citation data is often not available.

The effective use of bibliometrics has a number of requirements, not all of which are currently in place.

First, one needs robust and comprehensive data. As we have already seen, the main databases are reliable but their coverage is limited especially in the humanities and social sciences and they need to enlarge their scope to cover all forms of research outputs (Leydesdorff, 2008). Google Scholar is more comprehensive, but unreliable and non-transparent. At this time, full bibliometric evaluation is feasible in science and some areas of social science, but not in the humanities or some areas of technology (Archambault et al., 2006; Nederhof, 2006; van Leeuwen, 2006). Abramo and D'Angelo (2011)suggest that nations should routinely collect data on all the publications published within its institutions so that it is scrutinised and available on demand rather than having to be collected anew each time a research evaluation occurs (Ossenblok et al., 2012; Sivertsen & Larsen, 2012).

Second, one needs suitable metrics that measure what is important in as unbiased way as possible. These should not be crude ones such as simple counts of citations or papers, the h-index (although this has its advantages) or journal impact factors but more sophisticated ones that take into account the differences in citation practices across different disciplines as has been discussed in Section 4. This is currently an area of much debate with a range of possibilities (Gingras, 2014). The traditional crown indicator (now MNCS) is subject to criticisms concerning the use of the mean on highly cited data and also on the use of WoS field categories (Ruiz-Castillo & Waltman, 2014). There are source normalised alternatives such as SNIP or fractional counting (Aksnes et al., 2012) and metrics that include the prestige of the citing journals such as SJR. There are also moves towards non-parametric statistics based on percentiles. One dilemma is that the more sophisticated the metrics become, the less transparent and harder to replicate they are.

[25] http://www.shanghairanking.com/World-University-Rankings-2014/UK.html
[26] http://www.leidenranking.com/

A third area for consideration is inter- or trans- disciplinary work, and work that is more practical and practitioner oriented. How would this be affected by a move towards bibliometrics? There is currently little research in this area (Larivière & Gingras, 2010) although Rafols et al. (2012) found a systematic bias in research evaluation against interdisciplinary research in the field of business and management. Indeed, bibliometrics is still at the stage of establishing reliable and feasible methods for defining and measuring interdisciplinarity (Wagner et al., 2011). Huutoniemi et al. (2010) developed a typology and indicators to be applied to research proposals, and potentially research papers as well; Leydesdorff and Rafols (2011) have developed citation-based metrics to measure the interdisciplinarity of journals; and Silva et al. (Silva et al., 2013) evaluated the relative interdisciplinarity of science fields using entropy measures.

Fourth, we must recognise, and try to minimise, the fact that the act of measuring inevitably changes the behaviour of the people being measured. So, citation-based metrics will lead to practices, legitimate and illegitimate, to increase citations; an emphasis on 4* journals leads to a lack of innovation and a reinforcement of the status quo. For example, Moed (2008) detected significant patterns of response to UK research assessment metrics, with an increase in total publications after 1992 when numbers of papers were required; a shift to journals with higher citations after 1996 when quality was emphasised; and then in increase in the apparent number of research active staff through greater collaboration during 1997-2000. Michels and Schmoch (2014) found that German researchers changed their behaviour to aim for more US-based high impact journals in order to increase their citations.

Fifth, we must be aware that often problems are caused not by the data or metrics themselves, but by their inappropriate use either by academics or by administrators (Bornmann & Leydesdorff, 2014; van Raan, 2005b). There is often a desire for "quick and dirty" results and so simple measures such as the h-index or the JIF are used indiscriminately without due attention being paid to their limitations and biases. This also reminds us that there are ethical issues in the use of bibliometrics for research evaluation and Furner (2014) has developed a framework for evaluation that includes ethical dimensions.

## 8. Future Developments

### *8.1. Alternative metrics*

Although citations still form the core of scientometrics, the dramatic rise of social media has opened up many more channels for recording the impact of academic research (Bornmann, 2014; Konkiel & Scherer, 2013; Priem, 2014; Roemer & Borchardt, 2012). These go under the name of "altmetrics" both as a field, and as particular alternative metrics[27]. One of the interesting characteristics of altmetrics is that it throws light on the impacts of scholarly work on the general public rather than just the academic community. The Public Library of Science (PLoS) (Lin & Fenner, 2013) has produced a classification of types of impacts which goes from least significant to most significant:

- **Viewed**: institutional repositories, publishers, PLoS, Academia.com, ResearchGate. Perneger (2004) found a weak correlation with citations.
- **Downloaded/Saved**: as viewed plus CiteUlike, Mendelay.
- **Discussed**: Wikipedia, Facebook, Twitter, Natureblogs[28], ScienceSeeker[29], general research blogs. Eysenbach (2011) suggested a "twimpact factor" based on the number of tweets
- **Recommended**: F1000Prime[30]
- **Cited**: Wikipedia  CrossRef, WoS, Scopus, Google Scholar,

Altmetrics is still in its infancy and the majority of papers would have little social networking presence. There are also a number of problems: i) altmetrics can be gamed bv "buying" likes or tweets; ii) there is little by way of theory about how and why altmetrics are generated (this is also true of traditional citations); iii) a high score may not mean that the paper is especially good, just on a controversial or fashionable topic; and iv) because social media is relatively new it will under-represent older papers.

### *8.2. The shape of the discipline*

Citations refer to texts other than the one that contains the cited references, and thus induce a dynamic vision of the sciences developing as networks of relations (Price, 1965). In the scientometric literature, this has led to the call for "a theory of citation" (Cozzens, 1989; Cronin, 1998; Garfield, 1979; Leydesdorff, 1998; Nicolaisen, 2007). Wouters (1998) noted that in science and technology studies (STS), citations are studied as references in the context of "citing" practices, whereas the citation index inverts the directionality and studies "citedness" as a measure of impact. From the perspective of STS, the citation index thus would generate a semiotic artifact (Luukkonen, 1997).

[27] http://altmetrics.org/, http://www.altmetric.com/
[28] http://blogs.nature.com/
[29] http://scienceseeker.org/
[30] http://f1000.com/prime

References can have different functions in texts, such as legitimating research agendas, warranting knowledge claims, black-boxing discussions, or be perfunctory. In and among texts, references can also be compared with the co-occurrences and co-absences of words in a network model of science (Braam et al., 1991a; Braam et al., 1991b) A network theory of science was formulated by Hesse (1980, p. 83) as "an account that was first explicit in Duhem and more recently reinforced in Quine. Neither in Duhem nor in Quine, however, is it quite clear that the netlike interrelations between more observable predicates and their laws are in principle just as subject to modifications from the rest of the network as are those that are relatively theoretical." A network can be visualized, but can also be formalized as a matrix. The eigenvectors of the matrix span the latent dimensions of the network.

There is thus a bifurcation within the discipline of scientometrics. On the one hand, and by far the dominant partner, we have the relatively positivistic, quantitative analysis of citations as they have happened, after the fact so to speak. And on the other, we have the sociological, and often constructivist theorising about the generation of citations – a theory of citing behaviour. Clearly the two sides are, and need to be linked. The citing behaviour, as a set of generative mechanisms (Bhaskar, 1978), produces the citation events but, at the same time, analyses of the patterns of citations as "demi-regularities" (Lawson, 1997) can provide insights into the processes of scientific communication which can stimulate or validate theories of behaviour.

Another interesting approach is to consider the overall process as a social communication system. One could use Luhmann's (1995; 1996) theory of society as being based on autopoietic communication (Leydesdorff, 2007; Mingers, 2002). Different functional subsystems within society, e.g., science, generate their own organizationally closed networks of recursive communications. A communicative event consists of a unity of information, utterance and understanding between senders and receivers. Within the scientometrics context, the paper, its content and its publication would be the information and utterance, and the future references to it in other papers would be the understanding that it generates. Such communication systems operate at their own emergent level distinct from the individual scientists who underlie them, and generate their own cognitive distinctions that can be revealed the visualisation procedures discussed above.